\documentclass[twocolumn,secnumarabic,amssymb, nobibnotes, aps, prd]{revtex4-1}

\newcommand{\macro}[1]{\texttt{\textbackslash#1}}
\newcommand{\m}[1]{\macro{#1}}

\usepackage{hyperref}
\usepackage{amsmath}
\usepackage{multirow}
\usepackage{graphicx}
\usepackage[separate-uncertainty]{siunitx}[=v2]

\usepackage{soul}
\usepackage{color}

\begin{document}

\title{Determination of high-energy hadronic interaction properties from observables of proton initiated extensive air showers}%

\author{Isabel Astrid Goos}%
\affiliation{Laboratoire Astroparticule et Cosmologie (APC) - Paris \\
10 Rue Alice Domon et Léonie Duquet, 75013 Paris, France}
\author{Xavier Bertou}%
\affiliation{Centro Atómico Bariloche (CAB) - San Carlos de Bariloche\\
Av. Exequiel Bustillo 9500, 8400 San Carlos de Bariloche, Argentina}
\author{Tanguy Pierog}%
\affiliation{Karlsruher Institut für Technologie (KIT) - Karlsruhe\\
Hermann-von-Helmholtz-Platz 1, 76344 Eggenstein-Leopoldshafen, Germany}
\date{April 17, 2023}

\begin{abstract}
We propose a method to extract high-energy hadronic interaction properties from the distributions of two of the main observables of proton extensive air showers: the depth of maximum shower development, $X_\mathrm{max}$, and the number of muons at the ground, $N_\mu$. We determine relevant parameters of the first and subsequent interactions of the cascade and analyse how they impact on these observables. By training a universal neural network, we demonstrate that we can recover the most relevant parameters (fraction of energy going to the hadronic channel in the first interaction, first interaction multiplicity and effective inelasticity) for different hadronic interaction models using only the observables $X_\mathrm{max}$ and $N_\mu$.
\end{abstract}
\maketitle
\tableofcontents

\section{Introduction}

At their highest energies, cosmic rays arrive at Earth at an extremely low rate. As a consequence, they cannot be detected directly. Instead, they are observed via the extensive air showers (EAS) they produce upon interacting in the atmosphere \cite{engel2011extensive}. The first few interactions occur at energies above those accessible in man-made particle accelerators. An increasing statistic of ultra-high energy cosmic rays (UHECR, $\textrm{E} >$ \SI{e19}{\eV}) is being registered by large observatories such as the Pierre Auger Observatory \cite{pierre2015pierre} and the Telescope Array \cite{abu2012surface}. These hybrid observatories observe a fraction of the EAS, both through their longitudinal development in the atmosphere (employing fluorescence telescopes operating at night) and their lateral extension (using a ground array of particle detectors). In addition, the Pierre Auger Observatory is engaged in an upgrade phase, where more complementary detectors will allow for a multi-hybrid observation of EAS \cite{castellina2019augerprime}.

When an UHECR interacts in the atmosphere, it creates a core hadronic shower, mainly containing neutral and charged pions, which give rise to the electromagnetic and the muonic cascade, respectively \cite{engel2011extensive}. The electromagnetic cascade develops in the atmosphere and is then continuously absorbed, reaching a maximum development at a specific depth in the atmosphere, $X_\mathrm{max}$, that is related to the composition (and energy) of the primary cosmic ray. Muons travel from their production point to the ground only with little deflection and energy loss, compared to the electromagnetic cascade. The number of muons at the ground ($N_\mu$) is also related to the composition (and energy) of the primary cosmic ray. 
{Hybrid observations as done at the Pierre Auger Observatory and Telescope Array allow to extract both observables from EAS. Machine learning techniques \cite{aab2021deep, aab2021extraction} and detector upgrades \cite{castellina2019augerprime} significantly improve the extraction capabilities.

At a fixed energy, heavier nuclei tend to produce shallower EAS with more muons at the ground. Thus, an anticorrelation is present in measured distributions, that results from a mixed composition. However, even at a fixed composition and energy, an anticorrelation is expected between the shower maximum, $X_\mathrm{max}$, and the number of muons, $N_\mu$, which results from how energy is distributed between the electromagnetic and muonic cascades in the EAS. Figure~\ref{plot1} shows this anticorrelation for showers simulated with vertical proton primaries of \SI{e20}{\eV}.

\begin{figure}[h]
\centering
\includegraphics[width=.45\textwidth]{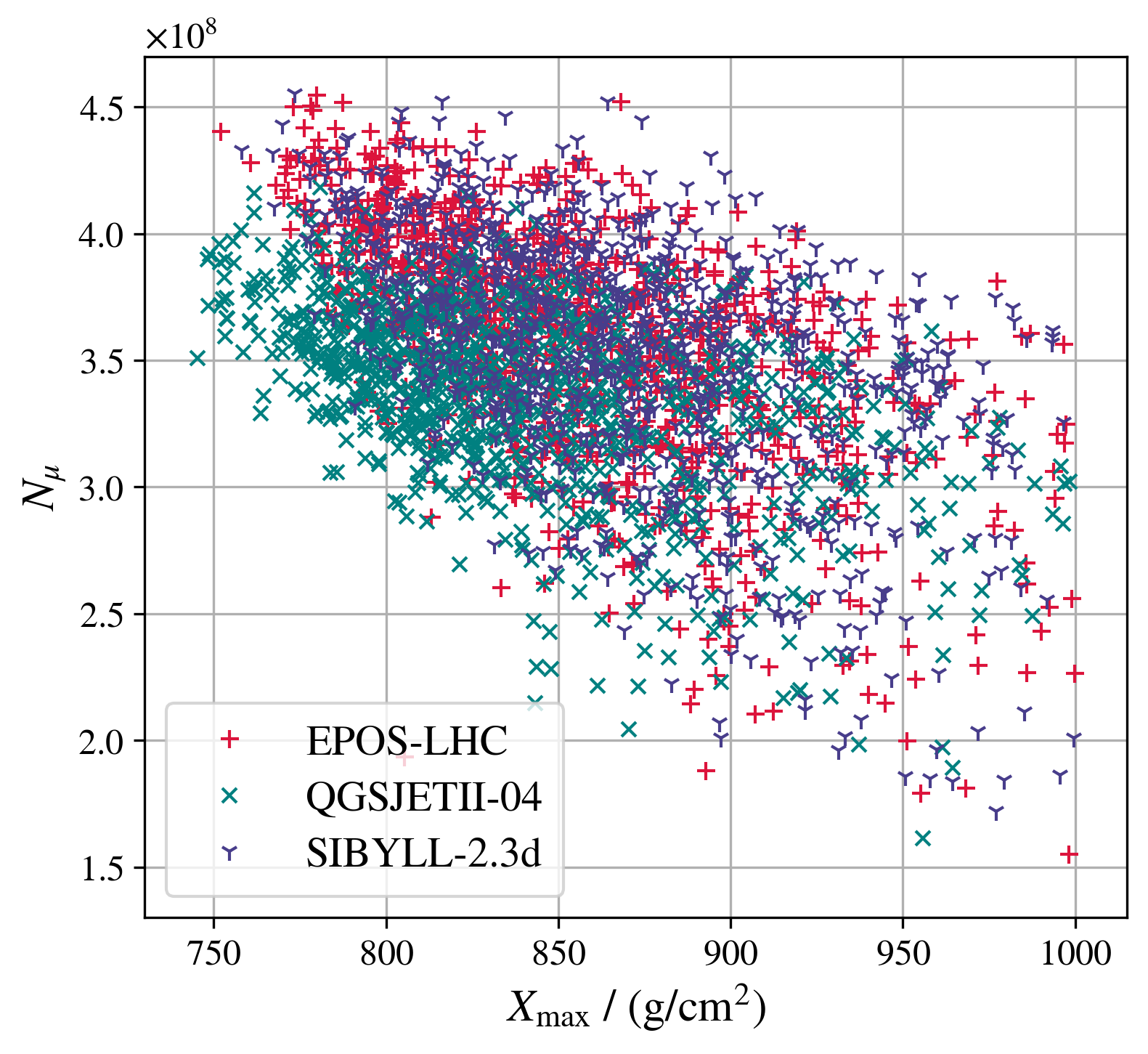}
\caption{Anticorrelation between the depth of maximum shower development, $X_\mathrm{max}$, and the number of muons at the ground, $N_\mu$. Three sets of 1000 showers induced by proton primaries of \SI{e20}{\eV} are shown. The sets were generated using the high-energy hadronic interaction models EPOS~LHC, QGSJETII-04 and Sybill~2.3d.}
\label{plot1}
\end{figure}

In order to properly understand and reconstruct the EAS of UHECR observed at Earth, it is necessary to resort to computer simulations. Different simulation codes are available, both for the generic EAS development (AIRES \cite{sciutto1999aires}, CORSIKA \cite{heck1998corsika}, CONEX \cite{Bergmann:2006yz}) and for the individual hadronic interactions, both at high (QGSJETII-04 \cite{Ostapchenko:2010gt}, EPOS~LHC\cite{Pierog:2013ria}, Sybill~2.3d \cite{Riehn:2019jet}, DPMJETIII\cite{Roesler:2000he}) and low energies (FLUKA \cite{Ferrari:2005zk}, GHEISHA \cite{jakubowski1990verified}, UrQMD \cite{Bleicher:1999xi}). The highest uncertainties in these simulations come from the high-energy hadronic interaction model implemented. On a shower to shower basis, considerable fluctuations are observed, which depend on the first few interactions, especially for light nuclei. When compared to observations, an overall muon deficit is observed in simulations \cite{aab2015muons, aab2021measurement, albrecht2022muon}, whichever high-energy hadronic interaction model is used when simulating. It is however difficult to correct models with current data, as there are multiple ways to modify the models in order to enhance the muon flux at the ground \cite{kampert2012measurements}.

While shower to shower fluctuations are an important feature when trying to determine the composition on a shower to shower basis, they can also be used to understand high-energy interactions. The $X_\mathrm{max}$ fluctuations can be used to estimate the proton-air cross-section at energies above those available in accelerators \cite{abreu2012measurement} or to pinpoint inconsistencies between data and models \cite{aab2021measurement, prado2019tests}. In this work, we study how the anticorrelation of two EAS observables, $X_\mathrm{max}$ and $N_\mu$, for a given mass, motivates a method to extract information on high-energy interactions. First, as will be described in section \ref{Search}, we search for parameters describing the physical processes in ultra-high energy proton-induced EAS that have a strong impact on $X_\mathrm{max}$ and $N_\mu$ (such as how energy is distributed in the first few interactions, what are the multiplicities of these interactions, etc.). Then, we train a neural network to model the $X_\mathrm{max}$ and $N_\mu$ observables as a function of these parameters. This process is described in section \ref{NN}. Once relevant parameters have been identified and a universal characterization of the $X_\mathrm{max}$-$N_\mu$ distribution is obtained (universal in the sense that only one neural network model is needed to reproduce the different current high-energy hadronic interaction scenarios simultaneously), we invert the system to determine the physical parameters for the different hadronic models based on their respective $X_\mathrm{max}$-$N_\mu$ distributions (see section \ref{IM}). This opens the way to devise a strategy to extract the physical parameters of observed cascades in the Pierre Auger Observatory or the Telescope Array and allows to estimate the systematic uncertainties involved (see section \ref{Res}).

\section{Search for physical parameters of extensive air showers}\label{Search}

The anticorrelation between $X_\mathrm{max}$ and $N_\mu$ (see figure~\ref{plot1}) can roughly and qualitatively be understood following an argument regarding the distribution of energy among product particles and along the shower. If a higher energy fraction is taken by hadronically interacting particles, then more muons can be created, while less energy is left to generate electromagnetic subshowers, which ultimately yields a lower value of $X_\mathrm{max}$ (and vice versa). However, the fact that the anticorrelation spreads considerably sideways along the slope means that there is some phenomenon adding another source of variability not captured by the argument just given. As mentioned in the introduction, previous works have focused on the $X_\mathrm{max}$ or $N_\mu$ distributions alone \cite{abreu2012measurement, cazon2021constraining}. The fact that these two observables are anticorrelated means that analyzing their joint distribution should render new information, not accessible when considering one of the marginal distributions alone. These arguments motivate a search for those physical parameters that have the strongest influence on $X_\mathrm{max}$ and $N_\mu$ simultaneously. In addition, knowing that the energy distribution among particles and the variability of it are at play, we have strong indications where to look for these parameters. In this section, we show that, for proton initiated showers, a considerable understanding of $X_\mathrm{max}$ and $N_\mu$ as a result of shower parameters is possible by including properties of the first interaction that impact on the energy distribution among particles and information on the leading particle in the semi-empirical model from Heitler \cite{heitler1948quantum} and Matthews \cite{matthews2005heitler}, which we briefly summarize next section (\ref{Extension}). 

\subsection{Extension of the Heitler model to proton initiated showers}\label{Extension}

In Matthews' extension \cite{matthews2005heitler} to Heitler's splitting approximation of electromagnetic cascades \cite{heitler1948quantum}, a primary proton is assumed to generate a set of $N_\mathrm{ch}$ charged and $N_0$ neutral pions when interacting with an atom in the atmosphere. Due to isospin invariance, $N_\mathrm{ch}$ is assumed to be twice the value of $N_0$. The charged pions interact, after travelling a characteristic interaction length $\lambda_\mathrm{I}$, yielding a new set of $N_\mathrm{ch}$ charged and $N_0$ neutral pions. On the other hand, all neutral pions created along the shower almost immediately decay to two photons that originate electromagnetic subshowers.

More specifically, one such photon sets off a sequence of pair creation and bremsstrahlung processes, in which the initial photon's energy is assumed to be distributed equally among particles. This progression stops as soon as individual energies drop below the electron critical energy $\xi_\mathrm{c}^\mathrm{e}$. From this point on, electrons are more likely to lose energy through collisions with molecules in the atmosphere than radiating photons, which leads to a decrease in the shower size. Thus, an electromagnetic subshower reaches a maximum number of particles after traversing an atmospheric burden of 
\begin{equation}\label{Xmaxgamma}
X_\mathrm{max}^\gamma = \lambda_\mathrm{r} \ln{\left( \frac{E_0^\gamma}{\xi_\mathrm{c}^\mathrm{e}} \right)},    
\end{equation}
where $E_0^\gamma$ is the energy of the initiating photon and $\lambda_\mathrm{r}$ is the radiation length of electromagnetic particles. 

The ensemble of all the electromagnetic subshowers created along the EAS dominates in number and energy the overall shower development. As a consequence, the depth $X_\mathrm{max}^\mathrm{p}$ at which the shower generated by a primary proton reaches its maximum number must depend heavily on the bulk of electromagnetic subshowers and, in particular, on the most influential ones. These are the ones initiated by the highest energy neutral pions, i.e. by the group of neutral pions that arise in the first interaction. Thus, expression (\ref{Xmaxgamma}) can be adapted to calculate the maximum depth of a proton shower as 
\begin{equation}\label{Xmaxp}
X_\mathrm{max}^\mathrm{p} = X_0 + \lambda_\mathrm{r} \ln{\left( \frac{E_0/(3 \times 2N_0)}{\xi_\mathrm{c}^\mathrm{e}} \right)}.    
\end{equation}
Here, $X_0$ is the depth where the first interaction occurs and $E_0/(3 \times 2N_0)$ represents the fact that in the first interaction one third of the primary energy is assumed to stay among the neutral pions which produce $2N_0$ photons.

Within the hadronic component, which in this approach consists of pions only, energy is assumed to be distributed evenly among an increasing number of neutral and charged pions. After $n_\mathrm{p}$ interaction steps, the latter reach the critical energy of the charged pions, $\xi_\mathrm{c}^\pi$. At this stage, it is more probable for them to decay than to interact. In Matthews' model, they are all assumed to decay to muons at this point. This simple model predicts a number 
\begin{equation}\label{Nmu}
N_\mu = \left( N_\mathrm{ch} \right) ^{n_\mathrm{p}}
\end{equation}
of muons from this point on until the ground level is reached. 

The number of interactions $n_\mathrm{p}$, necessary for the pions to reach their critical energy, can be calculated by equalizing the interaction length $\lambda_\mathrm{I}$ and the decay length $\lambda_\mathrm{dec} = \rho(h) \gamma c \tau_{\pi^\pm}$ of the pions \cite{kampert2012measurements}. Here, $\rho$ is the height-dependent density of air and $\gamma$ is the Lorentz factor of the pions when they reach their critical energy:
\begin{equation}\label{gamma}
\gamma = \frac{E_0/(N_0+N_\mathrm{ch})^{n_\mathrm{p}}}{m_{\pi^\pm}}.    
\end{equation}
If $\theta$ is the angle of incidence, $\lambda_\mathrm{I}$ can be obtained from 
\begin{equation}\label{costheta}
\cos{(\theta)} = \frac{\rho(h) h_\mathrm{s}}{n_\mathrm{p} \lambda_\mathrm{I}},
\end{equation}
where $h_\mathrm{s}$ is the mean scale height for the standard isothermal atmosphere. Inserting equations (\ref{gamma}) and (\ref{costheta}) in $\lambda_\mathrm{I} = \lambda_\mathrm{dec}$, one obtains the number of generations $n_\mathrm{p}$: 
\begin{equation}\label{np}
n_\mathrm{p} = - \frac{\mathrm{W}_{-1} \left( -\frac{h_\mathrm{s}}{c \tau_{\pi^\pm}} \frac{m_{\pi^\pm}}{E_0} \frac{\ln{(N_0+N_\mathrm{ch})}}{\cos{(\theta)}} \right)}{\ln{(N_0+N_\mathrm{ch})}}.
\end{equation}
$\mathrm{W_{-1}}$ denotes the lower branch of the Lambert-W function \cite{veberivc2012lambert}. Inserting this number of generations in expression (\ref{Nmu}), the number of muons can be obtained.

\subsection{Simulations}\label{Simus}

The expressions from section \ref{Extension} were formulated with the objective of describing average shower characteristics as a function of average physical parameters, i.e. without taking shower to shower fluctuations into account. Since our goal is to describe a distribution ($X_\mathrm{max}$-$N_\mu$ distribution) as a function of distributions of physical parameters, we need to verify if the expressions from section \ref{Extension} serve our purpose. We use simulations in order to evaluate these and to analyze where deficiencies emerge and where they might come from. This paves the way to devise an improvement of the semi-empirical model described in section \ref{Extension}, which will reveal the physical parameters of EAS we need in order to model the $X_\mathrm{max}$-$N_\mu$ distribution (see section \ref{IotSEM}). 

We will focus in this work on proton showers of \SI{e20}{\eV} because the $X_\mathrm{max}$-$N_\mu$ anticorrelation is more pronounced the higher the primary energy and the lower the primary mass are. Since our choice of physical parameters will be motivated by the processes that lead to an anticorrelation, it is suitable to carry out this first analysis in this setting. In order to produce a large enough set of simulations at this high energy, we use the simulation framework CONEX. It combines explicit Monte Carlo simulation of the highest energy portion of the air shower (first few interactions), with numerical expressions for the ensemble of low energy particles. Including the fluctuations introduced through the highest energy interactions ensures that one obtains accurate results for the shower to shower fluctuations in the EAS characteristics. Since the bulk of lower energy particles is large, particular characteristics are averaged out and there is no loss when dealing with these particles in a deterministic way using cascade equations \cite{kalmykov2003one}. Thus, average EAS parameters are reproduced accurately as well \cite{Bergmann:2006yz}. 

The treatment of hadronic interactions below and above a threshold energy of \SI{100}{\giga\eV} is handled by separate generators. For low energies we use URQMD, while for the high-energy end we employ EPOS~LHC, QGSJETII-04 and Sybill~2.3d. For each of these hadronic models, we simulate 1000 showers initiated by a vertically incident proton of \SI{e20}{\eV}. In these simulations performed with CONEX, we have access to the identity and energy of all the particles created in an interaction that was explicitly simulated by Monte Carlo. It is also possible to identify the set of particles created in the first interaction the primary proton undergoes, as well as the set of secondary particles produced in each later Monte Carlo generated interaction. With this information, we can calculate the physical parameters we analyze in this work. In addition, CONEX outputs longitudinal distributions from which we extract the values of $X_\mathrm{max}$ and $N_\mu$ corresponding to a particular shower. 

\subsection{Improvement of the semi-empirical model}\label{IotSEM}

While the assumption in Matthews' model that $N_\mathrm{ch} = 2 \times N_0$ in each interaction mostly holds, it is certainly not true that the energy is equally distributed among pions. This assumption leads to the factor of one third in expression (\ref{Xmaxp}). First, this neglects the fact that, when two hadrons interact, a significant fraction $(1-\kappa)$ of the available energy is carried by a single so-called leading particle. $\kappa$ is called the inelasticity and can considerably vary from shower to shower. For example, for our simulations, the inelasticity of the first interaction varies between 0.6 and 0.9 (not counting diffractive events, which have an inelasticity close to zero). Secondly, simulations reveal that there is a considerable variability in the fraction of energy that charged pions and other hadronically interacting particles carry. For the first interaction, this fraction spans values from $0.4$ to $0.8$. Thus, a first step in improving the modeling of $X_\mathrm{max}^\mathrm{p}$ is to replace the factor $1/3$ by a more realistic value of the energy fraction carried by hadronically interacting particles.

Following the idea that lead to expression (\ref{Xmaxp}), we calculate $X_\mathrm{max}^\mathrm{p}$ for the bulk of subshowers that come from the first interaction. The superscript ($\mathrm{FI}$) indicates that a parameter is specific to the first interaction. Taking into account the leading particle effect as well, we obtain the following expression: 
\begin{equation}\label{XmaxG}
X_\mathrm{max}^\mathrm{p} = X_0 + \lambda_\mathrm{r} \ln{\left( 
\frac{E_0 \: (1-f_\mathrm{ch}^\mathrm{FI}) \: \kappa^\mathrm{FI}}{\xi_\mathrm{c}^\mathrm{e} \: 2 \: (N_\mathrm{tot}^\mathrm{FI}/3)} \right)}.
\end{equation}
Here, the $(1/3)$ from expression (\ref{Xmaxp}) is replaced by $(1-f_\mathrm{ch}^\mathrm{FI}) \: \kappa^\mathrm{FI}$ to account for the fact that only a fraction $\kappa^\mathrm{FI}$ of the primary energy is at disposal for production of new particles, out of which $f_\mathrm{ch}^\mathrm{FI}$ is carried to later stages in the hadronic core. Additionally, $N_0$ is replaced by $(N_\mathrm{tot}^\mathrm{FI}/3)$. Finally, we can replace $2 \: (N_\mathrm{tot}^\mathrm{FI}/3)$ by $N_\mathrm{ch}^\mathrm{FI}$ because the relation $N_\mathrm{ch} = 2 \times N_0$, that involves multiplicities and not energies, approximately holds. 

For an improved calculation of the muon number, we propose to calculate the number of generations $n_\mathrm{p}$ as in (\ref{np}), but this time including the leading particle effect and separating the first interaction from the rest of the shower. This separation is crucial because the particle multiplicity and energy distribution among particles heavily depend on the available energy. In this sense, the first interaction not only stands out with a high variability in the parameters (as quantified in a previous section), but also with parameter values that differ significantly from those representative of the rest of the shower. For example, simulations reveal up to around 500 hadronically interacting particles with energy above $0.01 \%$ of the primary energy in the first interaction, while a representative number for the rest of the shower varies between around 11 and 13. We replace the numerator in expression (\ref{gamma}) by
\begin{equation}\label{socotroco1}
\gamma m_{\pi^\pm} = \frac{E_0 \: \left( 1- (1-f_\mathrm{ch}^\mathrm{FI}) \: \kappa^\mathrm{FI} \right) \: \left( 1- (1-f_\mathrm{ch}) \: \kappa \right)^{n_\mathrm{p}-1} }{(1+N_\mathrm{ch}^\mathrm{FI}) \: (1+N_\mathrm{ch})^{n_\mathrm{p}-1}}.    
\end{equation}
$\left( 1- (1-f_\mathrm{ch}^\mathrm{FI}) \: \kappa^\mathrm{FI} \right) \: \left( 1- (1-f_\mathrm{ch}) \: \kappa \right)^{n_\mathrm{p}-1}$ is the fraction of the primary energy that stays in the hadronic channel after $n_\mathrm{p}$ interactions. $((1-f_\mathrm{ch}) \: \kappa)$ is namely the energy that is diverted to the electromagnetic component. Then, we divide this expression for the energy by the number of hadronically interacting particles present after $n_\mathrm{p}$ interactions, separating the leading particle in each interaction from expressions $N_\mathrm{ch}$ and $N_\mathrm{ch}^\mathrm{FI}$.

Inserting expressions (\ref{socotroco1}) and (\ref{costheta}) in $\lambda_\mathrm{I} = \lambda_\mathrm{dec}$, one obtains
\begin{multline}\label{socotroco2}
n_\mathrm{p} = \mathrm{W}_{-1} \left( \ln{\left( \frac{1-(1-f_\mathrm{ch}) \: \kappa}{1+N_\mathrm{ch}} \right)} \cdot \frac{h_\mathrm{s}}{\cos(\theta)} \frac{m_{\pi^\pm} c^2}{c \tau_{\pi^\pm} E_0} \cdot \right. \\ \left. \frac{1+N_\mathrm{ch}^\mathrm{FI}}{1+N_\mathrm{ch}} \frac{1- (1-f_\mathrm{ch}) \: \kappa}{1- (1-f_\mathrm{ch}^\mathrm{FI}) \: \kappa^\mathrm{FI}} \right) \bigg/ \ln{\left( \frac{1-(1-f_\mathrm{ch}) \: \kappa}{1+N_\mathrm{ch}} \right)}.
\end{multline} 
Finally, the number of muons can be calculated as 
\begin{equation}\label{NmuG}
N_\mu = (1+N_\mathrm{ch}^\mathrm{FI}) \: (1+N_\mathrm{ch})^{n_\mathrm{p}-1}.
\end{equation}

\subsection{Performance of the improved model}

In order to assess the performance of our semi-empirical model, we need to define how to calculate the model parameters from our simulations. First, the multiplicities $N_\mathrm{ch}$ and $N_\mathrm{ch}^\mathrm{FI}$ and the fractions $f_\mathrm{ch}$ and $f_\mathrm{ch}^\mathrm{FI}$ need to be calculated taking into account that in every shower there is a considerable amount of particles other than pions. We assign to the multiplicities $N_\mathrm{ch}$ and $N_\mathrm{ch}^\mathrm{FI}$ all the particles that contribute to the muonic component: charged pions, baryons and kaons, leaving out the $\eta$ mesons in addition to the neutral pions (as in \cite{cazon2019probing}). We calculate the $N_\mathrm{ch}$ value of a shower as the geometric mean of all the individual $N_\mathrm{ch}^\mathrm{ind}$ values present in that shower (leaving out the first interaction). This assignment is motivated by the fact that multiplicities are multiplied in order to obtain meaningful expressions (see for example expression \ref{socotroco1}). For this same reason, we define $f_\mathrm{ch}$ in the same way. $N_\mathrm{ch}^\mathrm{FI}$ is the number of hadronically interacting particles in the first interaction with energy above $0.01 \%$ of the primary energy. Here, we leave low energy particles out because they have a very low impact on the shower development. $f_\mathrm{ch}^\mathrm{FI}$ of a shower is calculated as the average of all the fractions present in that shower, weighted by the energy available in each corresponding interaction. Thus, it is not strictly the value of the first interaction, but it is strongly correlated with it. More importantly, it implicitly corrects for the fact that in some showers the first interaction only amounts to some energy loss by the primary particle (diffractive events) and what happens in the second interaction is actually more important for the shower development. The inelasticity $\kappa^\mathrm{FI}$ is calculated as a weighted average as well, while $\kappa$ is taken to be the mode of the distribution of all the values happening in a shower. This choice is motivated by the fact that these distributions are strongly skewed.

Inserting, for each shower in each simulated set (EPOS~LHC, QGSJETII-04, Sybill~2.3d), the calculated model parameters into equations (\ref{XmaxG}) and (\ref{NmuG}), we obtain the $X_\mathrm{max}$ and $N_\mu$ values that all together yield an $X_\mathrm{max}$-$N_\mu$ distribution with mean values, standard deviations and correlation coefficient summarized in table \ref{table2}. We can compare these values with those directly obtained from the EPOS~LHC simulations, also summarized in table \ref{table2}. We observe that our semi-empirical model reproduces very well the mean values of the $X_\mathrm{max}$ and $N_\mu$ distributions. However, the calculated $X_\mathrm{max}$ distributions are a bit wider and the calculated anticorrelations are stronger than the original ones. For an analytical model, this performance is nevertheless very good. We conclude that the model parameters inserted into equations (\ref{XmaxG}) and (\ref{NmuG}) are good candidates to describe $X_\mathrm{max}$ and $N_\mu$.

\begin{table}
\begin{tabular}{ |c|c|c|c|c|c|c| } 
\hline
 & $\langle X_\mathrm{max} \rangle$ & $\sigma (X_\mathrm{max})$ & $\langle N_\mu \rangle$ & $\sigma (N_\mu)$ & $\rho(X_\mathrm{max}, N_\mu)$ \\
\hline
EPOS & \SI{821}{\g\per\centi\m\squared} & \SI{39}{\g\per\centi\m\squared} & 3.6e8 & 4.8e7 & -0.72\\ 
Calculations & \SI{822}{\g\per\centi\m\squared} & \SI{52}{\g\per\centi\m\squared} & 3.6e8 & 4.9e7 & -0.83\\ 
\hline
QGSJETII & \SI{800}{\g\per\centi\m\squared} & \SI{38}{\g\per\centi\m\squared} & 3.4e8 & 3.7e7 & -0.72\\ 
Calculations & \SI{813}{\g\per\centi\m\squared} & \SI{55}{\g\per\centi\m\squared} & 2.8e8 & 4.2e7 & -0.86\\ 
\hline
Sybill & \SI{827}{\g\per\centi\m\squared} & \SI{39}{\g\per\centi\m\squared} & 3.5e8 & 4.8e7 & -0.72\\ 
Calculations & \SI{817}{\g\per\centi\m\squared} & \SI{44}{\g\per\centi\m\squared} & 3.9e8 & 4.7e7 & -0.86\\ 
\hline

\end{tabular}
\caption{Mean values, standard deviations and correlation coefficient for the $X_\mathrm{max}$-$N_\mu$ distribution obtained using equations (\ref{XmaxG}) and (\ref{NmuG}) (second row in each case) and those values directly extracted from the corresponding simulated set (first row in each case).}
\label{table2}
\end{table}

Since our analytical model delivers a satisfactory prediction of the $X_\mathrm{max}$-$N_\mu$ distribution using the parameters described at the beginning of this section, we expect to be able to train a neural network on these features (called NN model from here on) in order to obtain a more refined NN model that predicts the targets $X_\mathrm{max}$ and $N_\mu$ more accurately. However, we require seven parameters in our model so far. In order to simplify it, we reduce the set of parameters down to the most influential ones. In this pursuit, we resort to random forest classifiers to calculate the feature importances using the Gini criterion. In addition to $X_0$, the three most important features are the three properties inherent to the first interaction, which is an interesting result by itself. This is not only valid for the prediction of the number of muons at the ground, which has already been discussed in the literature \cite{cazon2021constraining}, but also for the depth of maximum development. However, we need to keep uncorrelated parameters, because we will test our model with artificial parameter distributions, for which we need to be free from unknown constraints. The three most important and uncorrelated parameters are $\ln{(N_\mathrm{ch}^\mathrm{FI} + 1)}$, $f_\mathrm{ch}^\mathrm{FI}$ and $\kappa$. Here, we have replaced $N_\mathrm{ch}^\mathrm{FI}$ by $\ln{(N_\mathrm{ch}^\mathrm{FI} + 1)}$ because it is better practice to have more compactly distributed parameters as features in neural networks. We will use only these three parameters and $X_0$ as input features in the neural networks of the next section. 

\section{Method for the determination of high-energy hadronic interaction properties}

In this second step, we train neural networks in order to replace the analytical model described in section \ref{IotSEM}. Even though our analytical model is more complex than previous ones and demonstrates a very good performance, it is essentially based on successive discrete splittings, where energy is evenly distributed among particles belonging to the same group. It is to be expected that a neural network will identify other subtleties not captured by this simplified description of the processes involved. Finally, we will invert the system to deduce the physical parameters corresponding to the $X_\mathrm{max}$-$N_\mu$ distributions from the different simulation sets.

\subsection{Neural network modeling}\label{NN}

We decide to train on a dataset consisting of the three simulation sets described in section \ref{Simus} shuffled together (EPOS~LHC, QGSJETII-04 and Sybill~2.3d), which amounts to 3000 simulated showers in total. Since EAS simulations differ mostly at the highest energies, once those processes are captured in the form of physical parameters, the description of the rest of the shower should be common to all three high-energy interaction models considered here. Our objective is indeed to summarize the processes at the highest energies, which are not yet fully understood, in the form of parameters and model the rest of the shower with the help of neural networks. Using this mixed dataset, we expect to find a network that predicts $X_\mathrm{max}$ and $N_\mu$ in a model-independent way.

We separate $20 \%$ of the dataset to build our test set and $10 \%$ out of the remaining instances to build our validation set. Since our dataset is rather small, we carry out this separation following the stratified sampling technique, in order to have instances from less populated areas in the $X_\mathrm{max}$-$N_\mu$ distribution from which to learn. In order to predict $X_\mathrm{max}$ and $N_\mu$, the most important features are $\ln{(N_\mathrm{ch}^\mathrm{FI}+1)}$ and $f_\mathrm{ch}^\mathrm{FI}$, respectively. Since for each target observable it is justified to use a different set for this procedure of stratified sampling, we decide to develop a separate network for each of them, instead of aiming at a single model with a 2-dimensional output. We standardize the usual way by centering and scaling to unit variance each feature independently. As the $X_\mathrm{max}$-$N_\mu$ distribution has outliers, we decide to work with the mean absolute error as the loss function \cite{geron2022hands}.
 
For the prediction of $X_\mathrm{max}$, we obtain the best performance computing gradients on mini-batches of size 100 and updating the model's weights and biases using a learning rate of 1e-4. In order to avoid overfitting, we also employ momentum optimization with friction parameter 0.95, early stopping with a preset number of 10 epochs and $l_2$ regularization with a hyperparameter of $\alpha$=1e-5. The best results are obtained using the Relu function as the activation function in combination with the He initialization with a normal distribution for the definition of the initial weights and biases \cite{glorot2010understanding}. Finally, we use an architecture of 4 dense layers with 100 nodes each. For the prediction of $N_\mu$, we only need to change the batch size to 200, the learning rate to 5e-5, $\alpha$ to 5e-4 and the number of epochs to 25, in order to obtain the best results.

Using these configurations, we obtain neural networks that predict the values of $X_\mathrm{max}$ and $N_\mu$ as a function of $\ln{(N_\mathrm{ch}^\mathrm{FI} + 1)}$, $f_\mathrm{ch}^\mathrm{FI}$, $\kappa$ and $X_0$ very well. The absolute errors for the prediction of $X_\mathrm{max}$ on a shower to shower basis are close to \SI{30}{\gram\per\centi\metre\squared}, \SI{26}{\gram\per\centi\metre\squared} and \SI{30}{\gram\per\centi\metre\squared} for EPOS~LHC, QGSJETII-04 and Sybill~2.3d, respectively. On the other hand, the relative errors for the prediction of $N_\mu$ are of around $8.9 \%$, $6.0 \%$ and $8.3 \%$ for EPOS~LHC, QGSJETII-04 and Sybill~2.3d, respectively. A comparison between the original distributions from the simulations and those obtained from evaluating the neural networks at the shower parameters is shown in figure~\ref{plot2}, for the complete dataset but separated by hadronic interaction model. It reveals how the $X_\mathrm{max}$ and $N_\mu$ distributions differ between high-energy interaction models, but are still well captured by the corresponding neural network (both for $X_\mathrm{max}$ and $N_\mu$ prediction). We conclude that we have obtained a universal model.

\begin{figure*}
\centering
\begin{minipage}[b]{.4\textwidth}
\centering
\includegraphics[width=\textwidth]{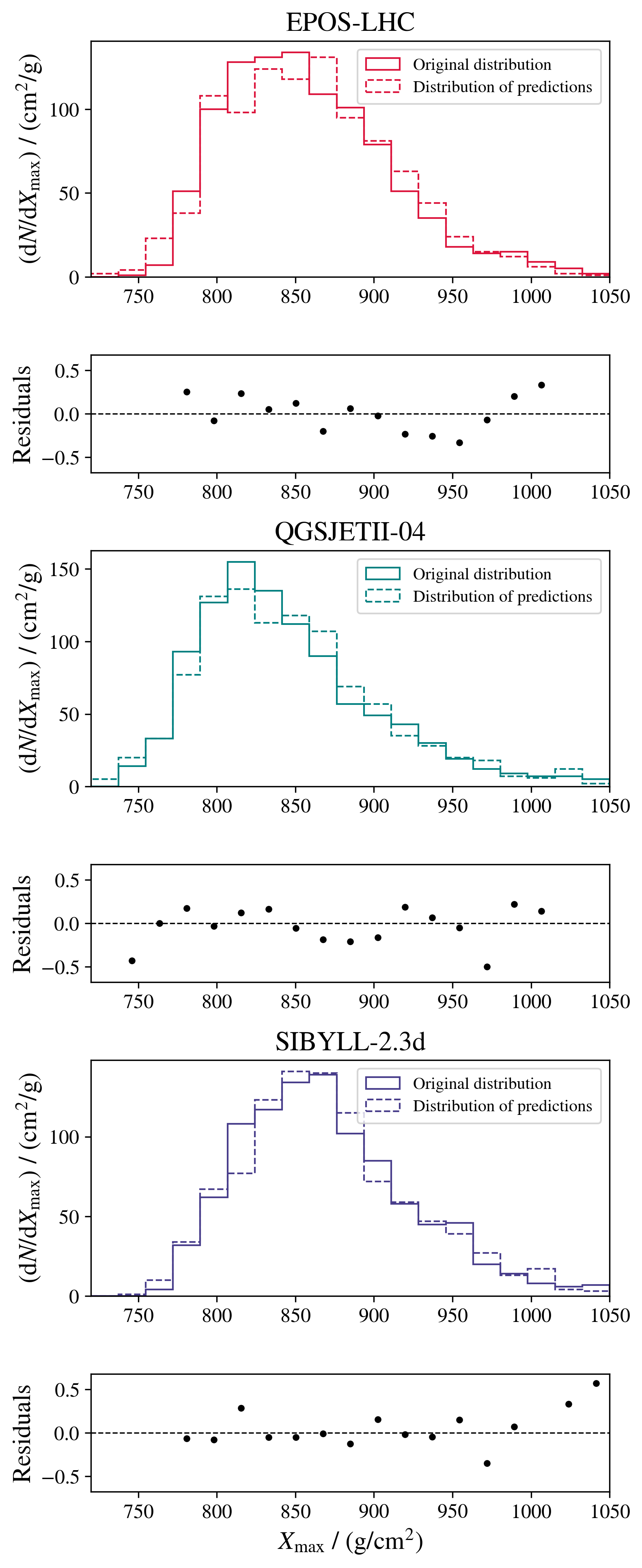}
\end{minipage}\qquad
\begin{minipage}[b]{.4\textwidth}
\centering
\includegraphics[width=\textwidth]{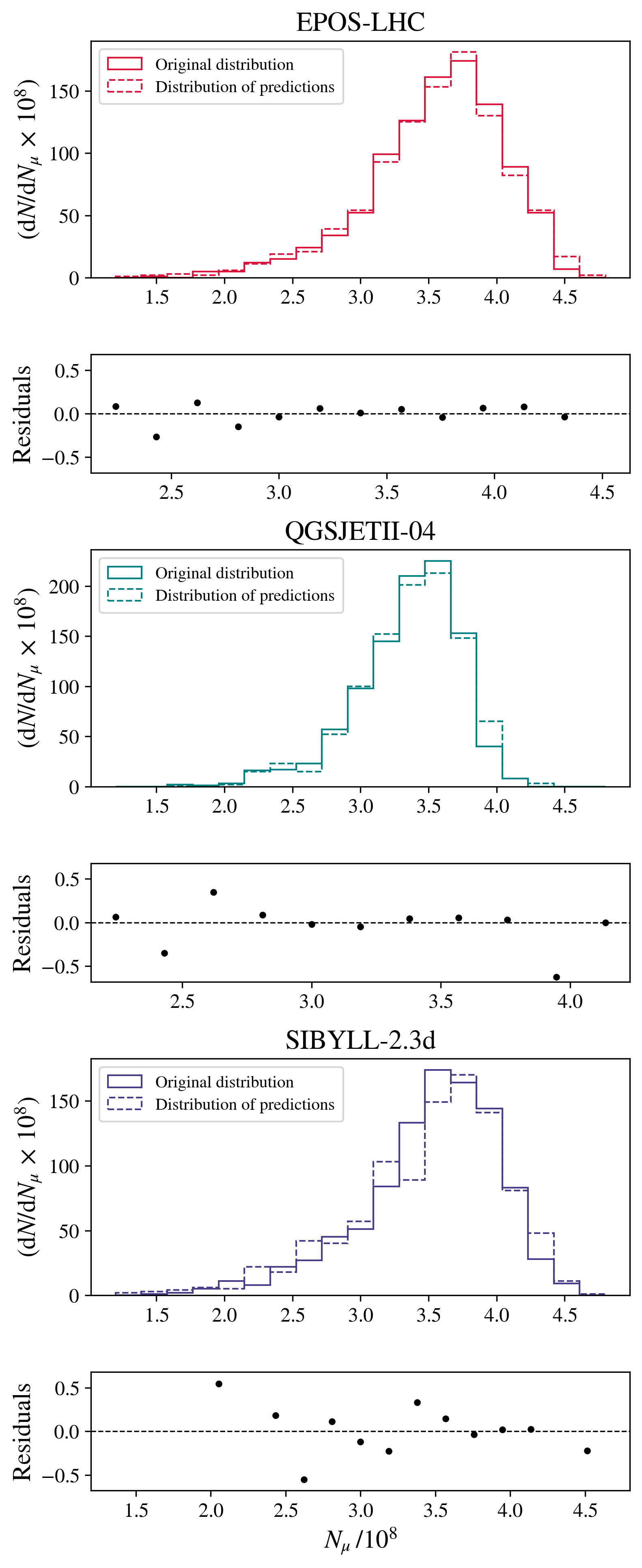}
\end{minipage}
\caption{Comparison between the original $X_\mathrm{max}$ (left) and $N_\mu$ (right) distributions from simulations (full lines) and those obtained from evaluating the NN model at the shower parameters of the complete dataset (dashed lines). We show the results for EPOS~LHC (top), QGSJETII-04 (middle) and Sybill~2.3d (bottom) separately. The $X_\mathrm{max}$ and $N_\mu$ distributions differ among high-energy interaction models, but are well captured by the corresponding neural network (both for $X_\mathrm{max}$ and $N_\mu$ prediction).}
\label{plot2}
\end{figure*}

\subsection{Inversion method}\label{IM}

In the previous section, we trained neural networks to predict $X_\mathrm{max}$ and $N_\mu$ as a function of specific values of $\ln{(N_\mathrm{ch}^\mathrm{FI} + 1)}$, $f_\mathrm{ch}^\mathrm{FI}$, $\kappa$ and $X_0$. We now switch to thinking of this designation as one between distributions. If we insert in our neural networks the distributions of the features, we obtain distributions of $X_\mathrm{max}$ and $N_\mu$ as an output. In order to work in this context, we need to parameterize the feature distributions. 

The distribution of $X_0$ is known \cite{abreu2012measurement} and thus the same for all three high-energy model scenarios. At $\SI{e20}{\eV}$, it can be described by the exponential distribution with a common value of $\lambda = \SI{40.4}{\g\per\centi\m\squared}$. $\ln{(N_\mathrm{ch}^\mathrm{FI} + 1)}$ and $f_\mathrm{ch}^\mathrm{FI}$ follow approximately a left-skewed Gumbel distribution, shifted according to a location parameter and rescaled according to a scale parameter. We define $N_\mathrm{loc}$ and $N_\mathrm{scale}$ as the location and scale parameters for $\ln{(N_\mathrm{ch}^\mathrm{FI} + 1)}$. $f_\mathrm{loc}$ and $f_\mathrm{scale}$ are equivalent values but for $f_\mathrm{ch}^\mathrm{FI}$. Finally, the $\kappa$ distributions can be approximated by normal distributions of mean $\kappa_\mathrm{loc}$ and standard deviation $\kappa_\mathrm{scale}$. The fitting parameter values we obtain for these distributions for each high-energy interaction model are summarized in table \ref{table1}. 

We notice that the values of $N_\mathrm{scale}$ and $\kappa_\mathrm{scale}$ coincide among all high-energy models. It is possible to see that the impact of changing these distribution parameters on the $X_\mathrm{max}$ and $N_\mu$ distributions is negligible \cite{Goos:2022dqv}. Consequently, we can consistently set their values to those shared among all three high-energy interaction models, without losing any generality and still having a procedure that does not depend on the hadronic model used. We are left with four to date unknown distribution parameters: $N_\mathrm{loc}$, $f_\mathrm{loc}$, $f_\mathrm{scale}$ and $\kappa_\mathrm{loc}$.

\begin{table}
\begin{tabular}{ |c|c|c|c|c|c|c|c| } 
\hline
 & $N_\mathrm{loc}$ & $N_\mathrm{scale}$ & $f_\mathrm{loc}$ & $f_\mathrm{scale}$ & $\kappa_\mathrm{loc}$ & $\kappa_\mathrm{scale}$ \\
\hline
EPOS~LHC & 6.153 & 1.05 & 0.720 & 0.062 & 0.738 & 0.013\\ 
QGSJETII-04 & 6.581 & 1.05 & 0.709 & 0.050 & 0.804 & 0.013\\ 
Sybill~2.3d & 6.055 & 1.05 & 0.723 & 0.061 & 0.710 & 0.013 \\ 
\hline
\end{tabular}
\caption{Fitting parameters for the distributions of $\ln{(N_\mathrm{ch}^\mathrm{FI} + 1)}$, $f_\mathrm{ch}^\mathrm{FI}$ and $\kappa$, as described in the text. These values depend on the high-energy hadronic interaction model considered.}
\label{table1}
\end{table}

We now proceed to describe how we invert the system. In summary, we perform a $\chi^2$-minimization where we compare the observables $\langle X_\mathrm{max} \rangle$, $\sigma ( X_\mathrm{max} )$, $\langle N_{\mu} \rangle$ and $\sigma ( N_{\mu} )$ obtained from a 4-dimensional grid of possible values for the parameters $N_\mathrm{loc}$, $f_\mathrm{loc}$, $f_\mathrm{scale}$ and $\kappa_\mathrm{loc}$ with the observables expected from EPOS~LHC, QGSJETII-04 and Sybill~2.3d simulations, in order to find those values in the grid that give the best agreement. The goal is to test the performance of our inversion method when predicting the unknown parameters, using the three scenarios we have available, and to evaluate if it can distinguish between the high-energy interaction models, thus revealing if our NN model is indeed universal. 

More concretely, for each hadronic interaction model $i$, we define
\begin{equation}
\chi_i^2(\bar{\theta}) = (\hat{\bar{x}}_i-\bar{\mu}) ^T V^{-1} (\hat{\bar{x}}_i-\bar{\mu}),
\label{eq:chi}
\end{equation}
where $\bar{\theta}$ represents an instance of the distribution parameters $N_\mathrm{loc}$, $f_\mathrm{loc}$, $f_\mathrm{scale}$ and $\kappa_\mathrm{loc}$. The vector $\bar{\mu}$ contains the values of the observables $\langle X_\mathrm{max} \rangle$, $\sigma ( X_\mathrm{max} )$, $\langle N_{\mu} \rangle$ and $\sigma ( N_{\mu} )$ obtained from generating $\ln{(N_\mathrm{ch}^\mathrm{FI} + 1)}$, $f_\mathrm{ch}^\mathrm{FI}$, $\kappa$ and $X_0$ distributions using the parameters $\bar{\theta}$ and evaluating the neural networks in these distributions. $\hat{\bar{x}}_i$ is an array containing the true values of the observables for the $i$-th high-energy scenario. We take the contributions due to statistical and systematic uncertainties into consideration: $$V = V_\mathrm{stat} + V_\mathrm{syst}. $$ Since the $X_\mathrm{max}$ and $N_{\mu}$ measurements are correlated, the covariance matrix $(V_\mathrm{stat})_{ij} = \mathrm{cov}[\hat{x}_i, \hat{x}_j]$ is used here \cite{lista2017statistical}, while for the systematic uncertainties we use $(V_\mathrm{syst})_{ij} = \mathrm{cov}[s_i, s_j]$, where ${s}$ contains the differences between the values of the observables for the $i$-th high-energy scenario and the values predicted when using the corresponding parameters (see table \ref{table1}). The entries for $\hat{x}_i$ and $s_i$, needed to compute $V_\mathrm{stat}$ and $V_\mathrm{syst}$, are estimated using the bootstrap method. The minimum of the $\chi_i^2$-function in equation (\ref{eq:chi}) defines the least-squares estimators $\hat{\bar{\theta}}_i$ we are looking for.

\subsection{Results}\label{Res}

In order to visualize the results, we work on 2-dimensional slices of the 4-dimensional parameter space. More precisely, for each high-energy scenario, we fix the values of $f_\mathrm{loc}$ and $f_\mathrm{scale}$ to where the minimum is obtained and fit the resulting 2-dimensional $\chi_i^2$ by a quadratic function, in order to find the corresponding estimators of $N_\mathrm{loc}$ and $\kappa_\mathrm{loc}$, together with the $1 \sigma$ (full lines) and $2 \sigma$ (dashed lines) confidence regions. The predictions for all high-energy interaction models are summarized in figure~\ref{plot3} (left). Equivalently, figure~\ref{plot3} (right) shows the predictions of $f_\mathrm{loc}$ and $f_\mathrm{scale}$ for the three high-energy interaction models. In both figures, the predicted values are marked with stars, while the true values for the corresponding high-energy scenario are indicated by filled circles.

\begin{figure*}
\centering
\begin{minipage}[b]{.45\textwidth}
\centering
\includegraphics[width=\textwidth]{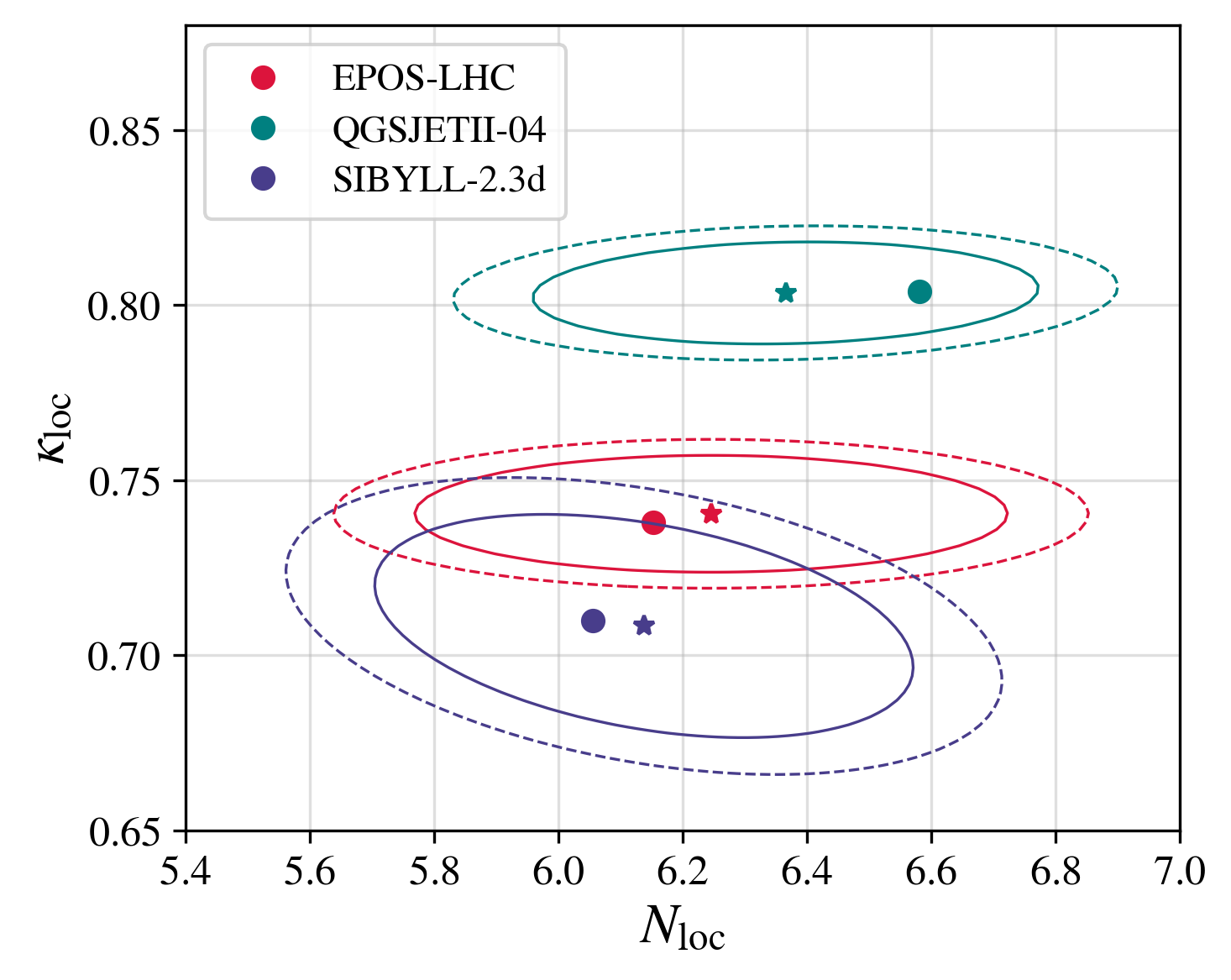}
\end{minipage}\qquad
\begin{minipage}[b]{.45\textwidth}
\centering
\includegraphics[width=\textwidth]{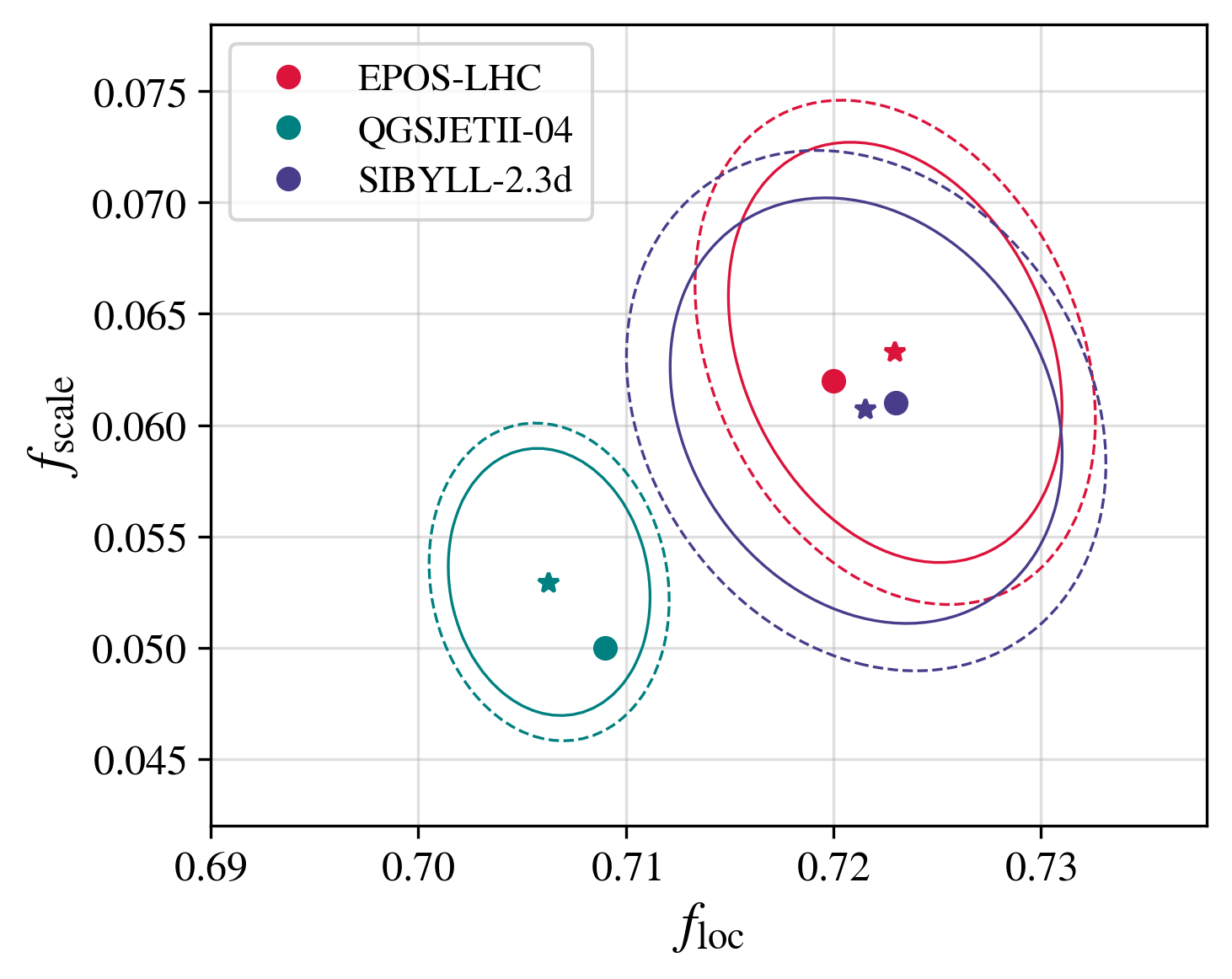}
\end{minipage}
\caption{Predictions (stars) of $N_\mathrm{loc}$ and $\kappa_\mathrm{loc}$ (left) and of $f_\mathrm{loc}$ and $f_\mathrm{scale}$ (right), as described in the text. Also shown are the $1 \sigma$ (full lines) and $2 \sigma$ (dashed lines) confidence regions for these predictions and the original values for each model (filled circles).}
\label{plot3}
\end{figure*}

Figure~\ref{plot3} reveals that the predictions of $N_\mathrm{loc}$, $\kappa_\mathrm{loc}$, $f_\mathrm{loc}$ and $f_\mathrm{scale}$ are well within $1 \sigma$ of the true values. In addition, in most of the cases, the prediction for one hadronic model is outside the $2 \sigma$ region of the other models. Only the $1 \sigma$ confidence region for the predictions of $f_\mathrm{loc}$ and $f_\mathrm{scale}$ for EPOS~LHC and Sybill~2.3d overlap considerably. Nevertheless, the predicted values exhibit similar differences among them compared to the true values. Furthermore, these parameters were very close to begin with. We conclude that we have found a universal model and a method to invert it. Additionally, the errors obtained here can be used as a measure of systematic uncertainties in a future application to a set of data.

\section{Conclusions}

Motivated by the anticorrelation between $X_\mathrm{max}$ and $N_\mu$ of \SI{e20}{\eV} proton showers, we developed an analytical model that reproduces this distribution as a function of parameters describing the multiplicity of hadronically interacting particles, the fraction of energy that is taken by these particles and the inelasticity of the first interaction and corresponding effective macro-parameters representative of the whole shower. We then replaced the analytical model by a neural network model. In this step, we identified the minimal set of physical parameters of EAS that are essential to understand the $X_\mathrm{max}$-$N_\mu$ distribution in a model-independent way: $\ln{(N_\mathrm{ch}^\mathrm{FI} + 1)}$, $f_\mathrm{ch}^\mathrm{FI}$, $\kappa$ and $X_0$. We concluded outlining and validating an inversion method that can be applied in the future to a measured $X_\mathrm{max}$-$N_\mu$ distribution in order to obtain the physical distributions, which themselves will impose new constraints on hadronic interaction models. 

We observed that the performance of the neural network model is remarkably good, especially considering that we are summarizing whole complex showers in only four parameters and a neural network model. The discrepancies we observe come from the fact that we are fitting the parameter distributions and that the neural network model is trained on a minimal set of features. This result opens the door for the development of similar models using showers of lower energies, where most of the statistics of the current giant observatories lies. 

\begin{acknowledgments}
Isabel Goos has been supported by the Consejo Nacional de Investigaciones Científicas y Técnicas (CONICET), the German Argentinean University Center (CUAA-DAHZ) and the Helmholtz International Research School for Astroparticle Physics and Enabling Technologies (HIRSAP).
\end{acknowledgments}

\bibliography{apsguide4-2}

\begin{thebibliography}{32}%
\makeatletter
\providecommand \@ifxundefined [1]{%
 \@ifx{#1\undefined}
}%
\providecommand \@ifnum [1]{%
 \ifnum #1\expandafter \@firstoftwo
 \else \expandafter \@secondoftwo
 \fi
}%
\providecommand \@ifx [1]{%
 \ifx #1\expandafter \@firstoftwo
 \else \expandafter \@secondoftwo
 \fi
}%
\providecommand \natexlab [1]{#1}%
\providecommand \enquote  [1]{``#1''}%
\providecommand \bibnamefont  [1]{#1}%
\providecommand \bibfnamefont [1]{#1}%
\providecommand \citenamefont [1]{#1}%
\providecommand \href@noop [0]{\@secondoftwo}%
\providecommand \href [0]{\begingroup \@sanitize@url \@href}%
\providecommand \@href[1]{\@@startlink{#1}\@@href}%
\providecommand \@@href[1]{\endgroup#1\@@endlink}%
\providecommand \@sanitize@url [0]{\catcode `\\12\catcode `\$12\catcode
  `\&12\catcode `\#12\catcode `\^12\catcode `\_12\catcode `\%12\relax}%
\providecommand \@@startlink[1]{}%
\providecommand \@@endlink[0]{}%
\providecommand \url  [0]{\begingroup\@sanitize@url \@url }%
\providecommand \@url [1]{\endgroup\@href {#1}{\urlprefix }}%
\providecommand \urlprefix  [0]{URL }%
\providecommand \Eprint [0]{\href }%
\providecommand \doibase [0]{http://dx.doi.org/}%
\providecommand \selectlanguage [0]{\@gobble}%
\providecommand \bibinfo  [0]{\@secondoftwo}%
\providecommand \bibfield  [0]{\@secondoftwo}%
\providecommand \translation [1]{[#1]}%
\providecommand \BibitemOpen [0]{}%
\providecommand \bibitemStop [0]{}%
\providecommand \bibitemNoStop [0]{.\EOS\space}%
\providecommand \EOS [0]{\spacefactor3000\relax}%
\providecommand \BibitemShut  [1]{\csname bibitem#1\endcsname}%
\let\auto@bib@innerbib\@empty
\bibitem [{\citenamefont {Engel}\ \emph {et~al.}(2011)\citenamefont {Engel},
  \citenamefont {Heck},\ and\ \citenamefont {Pierog}}]{engel2011extensive}%
  \BibitemOpen
  \bibfield  {author} {\bibinfo {author} {\bibfnamefont {R.}~\bibnamefont
  {Engel}}, \bibinfo {author} {\bibfnamefont {D.}~\bibnamefont {Heck}}, \ and\
  \bibinfo {author} {\bibfnamefont {T.}~\bibnamefont {Pierog}},\ }\href@noop {}
  {\bibfield  {journal} {\bibinfo  {journal} {Annual review of nuclear and
  particle science}\ }\textbf {\bibinfo {volume} {61}},\ \bibinfo {pages} {467}
  (\bibinfo {year} {2011})}\BibitemShut {NoStop}%
\bibitem [{\citenamefont {Collaboration}\ \emph {et~al.}(2015)\citenamefont
  {Collaboration} \emph {et~al.}}]{pierre2015pierre}%
  \BibitemOpen
  \bibfield  {author} {\bibinfo {author} {\bibfnamefont {P.~A.}\ \bibnamefont
  {Collaboration}} \emph {et~al.},\ }\href@noop {} {\bibfield  {journal}
  {\bibinfo  {journal} {Nuclear Instruments and Methods in Physics Research
  Section A: Accelerators, Spectrometers, Detectors and Associated Equipment}\
  }\textbf {\bibinfo {volume} {798}},\ \bibinfo {pages} {172} (\bibinfo {year}
  {2015})}\BibitemShut {NoStop}%
\bibitem [{\citenamefont {Abu-Zayyad}\ \emph {et~al.}(2012)\citenamefont
  {Abu-Zayyad}, \citenamefont {Aida}, \citenamefont {Allen}, \citenamefont
  {Anderson}, \citenamefont {Azuma}, \citenamefont {Barcikowski}, \citenamefont
  {Belz}, \citenamefont {Bergman}, \citenamefont {Blake}, \citenamefont {Cady}
  \emph {et~al.}}]{abu2012surface}%
  \BibitemOpen
  \bibfield  {author} {\bibinfo {author} {\bibfnamefont {T.}~\bibnamefont
  {Abu-Zayyad}}, \bibinfo {author} {\bibfnamefont {R.}~\bibnamefont {Aida}},
  \bibinfo {author} {\bibfnamefont {M.}~\bibnamefont {Allen}}, \bibinfo
  {author} {\bibfnamefont {R.}~\bibnamefont {Anderson}}, \bibinfo {author}
  {\bibfnamefont {R.}~\bibnamefont {Azuma}}, \bibinfo {author} {\bibfnamefont
  {E.}~\bibnamefont {Barcikowski}}, \bibinfo {author} {\bibfnamefont
  {J.}~\bibnamefont {Belz}}, \bibinfo {author} {\bibfnamefont {D.}~\bibnamefont
  {Bergman}}, \bibinfo {author} {\bibfnamefont {S.}~\bibnamefont {Blake}},
  \bibinfo {author} {\bibfnamefont {R.}~\bibnamefont {Cady}},  \emph {et~al.},\
  }\href@noop {} {\bibfield  {journal} {\bibinfo  {journal} {Nuclear
  Instruments and Methods in Physics Research Section A: Accelerators,
  Spectrometers, Detectors and Associated Equipment}\ }\textbf {\bibinfo
  {volume} {689}},\ \bibinfo {pages} {87} (\bibinfo {year} {2012})}\BibitemShut
  {NoStop}%
\bibitem [{\citenamefont {Castellina}(2019)}]{castellina2019augerprime}%
  \BibitemOpen
  \bibfield  {author} {\bibinfo {author} {\bibfnamefont {A.}~\bibnamefont
  {Castellina}},\ }in\ \href@noop {} {\emph {\bibinfo {booktitle} {EPJ Web of
  Conferences}}},\ Vol.\ \bibinfo {volume} {210}\ (\bibinfo {organization} {EDP
  Sciences},\ \bibinfo {year} {2019})\ p.\ \bibinfo {pages} {06002}\BibitemShut
  {NoStop}%
\bibitem [{\citenamefont {Aab}\ \emph {et~al.}(2021{\natexlab{a}})\citenamefont
  {Aab}, \citenamefont {Abreu}, \citenamefont {Aglietta}, \citenamefont
  {Albury}, \citenamefont {Allekotte}, \citenamefont {Almela}, \citenamefont
  {Alvarez-Mu{\~n}iz}, \citenamefont {Batista}, \citenamefont {Anastasi},
  \citenamefont {Anchordoqui} \emph {et~al.}}]{aab2021deep}%
  \BibitemOpen
  \bibfield  {author} {\bibinfo {author} {\bibfnamefont {A.}~\bibnamefont
  {Aab}}, \bibinfo {author} {\bibfnamefont {P.}~\bibnamefont {Abreu}}, \bibinfo
  {author} {\bibfnamefont {M.}~\bibnamefont {Aglietta}}, \bibinfo {author}
  {\bibfnamefont {J.~M.}\ \bibnamefont {Albury}}, \bibinfo {author}
  {\bibfnamefont {I.}~\bibnamefont {Allekotte}}, \bibinfo {author}
  {\bibfnamefont {A.}~\bibnamefont {Almela}}, \bibinfo {author} {\bibfnamefont
  {J.}~\bibnamefont {Alvarez-Mu{\~n}iz}}, \bibinfo {author} {\bibfnamefont
  {R.~A.}\ \bibnamefont {Batista}}, \bibinfo {author} {\bibfnamefont {G.~A.}\
  \bibnamefont {Anastasi}}, \bibinfo {author} {\bibfnamefont {L.}~\bibnamefont
  {Anchordoqui}},  \emph {et~al.},\ }\href@noop {} {\bibfield  {journal}
  {\bibinfo  {journal} {Journal of instrumentation}\ }\textbf {\bibinfo
  {volume} {16}},\ \bibinfo {pages} {P07019} (\bibinfo {year}
  {2021}{\natexlab{a}})}\BibitemShut {NoStop}%
\bibitem [{\citenamefont {Aab}\ \emph {et~al.}(2021{\natexlab{b}})\citenamefont
  {Aab}, \citenamefont {Abreu}, \citenamefont {Aglietta}, \citenamefont
  {Albury}, \citenamefont {Allekotte}, \citenamefont {Almela}, \citenamefont
  {Alvarez-Mu{\~n}iz}, \citenamefont {Batista}, \citenamefont {Anastasi},
  \citenamefont {Anchordoqui} \emph {et~al.}}]{aab2021extraction}%
  \BibitemOpen
  \bibfield  {author} {\bibinfo {author} {\bibfnamefont {A.}~\bibnamefont
  {Aab}}, \bibinfo {author} {\bibfnamefont {P.}~\bibnamefont {Abreu}}, \bibinfo
  {author} {\bibfnamefont {M.}~\bibnamefont {Aglietta}}, \bibinfo {author}
  {\bibfnamefont {J.}~\bibnamefont {Albury}}, \bibinfo {author} {\bibfnamefont
  {I.}~\bibnamefont {Allekotte}}, \bibinfo {author} {\bibfnamefont
  {A.}~\bibnamefont {Almela}}, \bibinfo {author} {\bibfnamefont
  {J.}~\bibnamefont {Alvarez-Mu{\~n}iz}}, \bibinfo {author} {\bibfnamefont
  {R.~A.}\ \bibnamefont {Batista}}, \bibinfo {author} {\bibfnamefont
  {G.}~\bibnamefont {Anastasi}}, \bibinfo {author} {\bibfnamefont
  {L.}~\bibnamefont {Anchordoqui}},  \emph {et~al.},\ }\href@noop {} {\bibfield
   {journal} {\bibinfo  {journal} {Journal of instrumentation}\ }\textbf
  {\bibinfo {volume} {16}},\ \bibinfo {pages} {P07016} (\bibinfo {year}
  {2021}{\natexlab{b}})}\BibitemShut {NoStop}%
\bibitem [{\citenamefont {Sciutto}(1999)}]{sciutto1999aires}%
  \BibitemOpen
  \bibfield  {author} {\bibinfo {author} {\bibfnamefont {S.}~\bibnamefont
  {Sciutto}},\ }\href@noop {} {\bibfield  {journal} {\bibinfo  {journal} {arXiv
  preprint astro-ph/9911331}\ } (\bibinfo {year} {1999})}\BibitemShut {NoStop}%
\bibitem [{\citenamefont {Heck}\ \emph {et~al.}(1998)\citenamefont {Heck},
  \citenamefont {Knapp}, \citenamefont {Capdevielle}, \citenamefont {Schatz},
  \citenamefont {Thouw} \emph {et~al.}}]{heck1998corsika}%
  \BibitemOpen
  \bibfield  {author} {\bibinfo {author} {\bibfnamefont {D.}~\bibnamefont
  {Heck}}, \bibinfo {author} {\bibfnamefont {J.}~\bibnamefont {Knapp}},
  \bibinfo {author} {\bibfnamefont {J.}~\bibnamefont {Capdevielle}}, \bibinfo
  {author} {\bibfnamefont {G.}~\bibnamefont {Schatz}}, \bibinfo {author}
  {\bibfnamefont {T.}~\bibnamefont {Thouw}},  \emph {et~al.},\ }\href@noop {}
  {\bibfield  {journal} {\bibinfo  {journal} {Report fzka}\ }\textbf {\bibinfo
  {volume} {6019}} (\bibinfo {year} {1998})}\BibitemShut {NoStop}%
\bibitem [{\citenamefont {Bergmann}\ \emph {et~al.}(2007)\citenamefont
  {Bergmann}, \citenamefont {Engel}, \citenamefont {Heck}, \citenamefont
  {Kalmykov}, \citenamefont {Ostapchenko}, \citenamefont {Pierog},
  \citenamefont {Thouw},\ and\ \citenamefont {Werner}}]{Bergmann:2006yz}%
  \BibitemOpen
  \bibfield  {author} {\bibinfo {author} {\bibfnamefont {T.}~\bibnamefont
  {Bergmann}}, \bibinfo {author} {\bibfnamefont {R.}~\bibnamefont {Engel}},
  \bibinfo {author} {\bibfnamefont {D.}~\bibnamefont {Heck}}, \bibinfo {author}
  {\bibfnamefont {N.~N.}\ \bibnamefont {Kalmykov}}, \bibinfo {author}
  {\bibfnamefont {S.}~\bibnamefont {Ostapchenko}}, \bibinfo {author}
  {\bibfnamefont {T.}~\bibnamefont {Pierog}}, \bibinfo {author} {\bibfnamefont
  {T.}~\bibnamefont {Thouw}}, \ and\ \bibinfo {author} {\bibfnamefont
  {K.}~\bibnamefont {Werner}},\ }\href {\doibase
  10.1016/j.astropartphys.2006.08.005} {\bibfield  {journal} {\bibinfo
  {journal} {Astropart. Phys.}\ }\textbf {\bibinfo {volume} {26}},\ \bibinfo
  {pages} {420} (\bibinfo {year} {2007})},\ \Eprint
  {http://arxiv.org/abs/astro-ph/0606564} {arXiv:astro-ph/0606564} \BibitemShut
  {NoStop}%
\bibitem [{\citenamefont {Ostapchenko}(2010)}]{Ostapchenko:2010gt}%
  \BibitemOpen
  \bibfield  {author} {\bibinfo {author} {\bibfnamefont {S.}~\bibnamefont
  {Ostapchenko}},\ }\href {\doibase 10.1103/PhysRevD.81.114028} {\bibfield
  {journal} {\bibinfo  {journal} {Phys. Rev. D}\ }\textbf {\bibinfo {volume}
  {81}},\ \bibinfo {pages} {114028} (\bibinfo {year} {2010})},\ \Eprint
  {http://arxiv.org/abs/1003.0196} {arXiv:1003.0196 [hep-ph]} \BibitemShut
  {NoStop}%
\bibitem [{\citenamefont {Pierog}\ \emph {et~al.}(2015)\citenamefont {Pierog},
  \citenamefont {Karpenko}, \citenamefont {Katzy}, \citenamefont {Yatsenko},\
  and\ \citenamefont {Werner}}]{Pierog:2013ria}%
  \BibitemOpen
  \bibfield  {author} {\bibinfo {author} {\bibfnamefont {T.}~\bibnamefont
  {Pierog}}, \bibinfo {author} {\bibfnamefont {I.}~\bibnamefont {Karpenko}},
  \bibinfo {author} {\bibfnamefont {J.~M.}\ \bibnamefont {Katzy}}, \bibinfo
  {author} {\bibfnamefont {E.}~\bibnamefont {Yatsenko}}, \ and\ \bibinfo
  {author} {\bibfnamefont {K.}~\bibnamefont {Werner}},\ }\href {\doibase
  10.1103/PhysRevC.92.034906} {\bibfield  {journal} {\bibinfo  {journal} {Phys.
  Rev. C}\ }\textbf {\bibinfo {volume} {92}},\ \bibinfo {pages} {034906}
  (\bibinfo {year} {2015})},\ \Eprint {http://arxiv.org/abs/1306.0121}
  {arXiv:1306.0121 [hep-ph]} \BibitemShut {NoStop}%
\bibitem [{\citenamefont {Riehn}\ \emph {et~al.}(2020)\citenamefont {Riehn},
  \citenamefont {Engel}, \citenamefont {Fedynitch}, \citenamefont {Gaisser},\
  and\ \citenamefont {Stanev}}]{Riehn:2019jet}%
  \BibitemOpen
  \bibfield  {author} {\bibinfo {author} {\bibfnamefont {F.}~\bibnamefont
  {Riehn}}, \bibinfo {author} {\bibfnamefont {R.}~\bibnamefont {Engel}},
  \bibinfo {author} {\bibfnamefont {A.}~\bibnamefont {Fedynitch}}, \bibinfo
  {author} {\bibfnamefont {T.~K.}\ \bibnamefont {Gaisser}}, \ and\ \bibinfo
  {author} {\bibfnamefont {T.}~\bibnamefont {Stanev}},\ }\href {\doibase
  10.1103/PhysRevD.102.063002} {\bibfield  {journal} {\bibinfo  {journal}
  {Phys. Rev. D}\ }\textbf {\bibinfo {volume} {102}},\ \bibinfo {pages}
  {063002} (\bibinfo {year} {2020})},\ \Eprint
  {http://arxiv.org/abs/1912.03300} {arXiv:1912.03300 [hep-ph]} \BibitemShut
  {NoStop}%
\bibitem [{\citenamefont {Roesler}\ \emph {et~al.}(2000)\citenamefont
  {Roesler}, \citenamefont {Engel},\ and\ \citenamefont
  {Ranft}}]{Roesler:2000he}%
  \BibitemOpen
  \bibfield  {author} {\bibinfo {author} {\bibfnamefont {S.}~\bibnamefont
  {Roesler}}, \bibinfo {author} {\bibfnamefont {R.}~\bibnamefont {Engel}}, \
  and\ \bibinfo {author} {\bibfnamefont {J.}~\bibnamefont {Ranft}},\ }in\ \href
  {\doibase 10.1007/978-3-642-18211-2_166} {\emph {\bibinfo {booktitle}
  {{International Conference on Advanced Monte Carlo for Radiation Physics,
  Particle Transport Simulation and Applications (MC 2000)}}}}\ (\bibinfo
  {year} {2000})\ pp.\ \bibinfo {pages} {1033--1038},\ \Eprint
  {http://arxiv.org/abs/hep-ph/0012252} {arXiv:hep-ph/0012252} \BibitemShut
  {NoStop}%
\bibitem [{\citenamefont {Ferrari}\ \emph {et~al.}(2005)\citenamefont
  {Ferrari}, \citenamefont {Sala}, \citenamefont {Fasso},\ and\ \citenamefont
  {Ranft}}]{Ferrari:2005zk}%
  \BibitemOpen
  \bibfield  {author} {\bibinfo {author} {\bibfnamefont {A.}~\bibnamefont
  {Ferrari}}, \bibinfo {author} {\bibfnamefont {P.~R.}\ \bibnamefont {Sala}},
  \bibinfo {author} {\bibfnamefont {A.}~\bibnamefont {Fasso}}, \ and\ \bibinfo
  {author} {\bibfnamefont {J.}~\bibnamefont {Ranft}},\ }\href {\doibase
  10.2172/877507} {\  (\bibinfo {year} {2005}),\ 10.2172/877507}\BibitemShut
  {NoStop}%
\bibitem [{\citenamefont {Jakubowski}\ and\ \citenamefont
  {Kobel}(1990)}]{jakubowski1990verified}%
  \BibitemOpen
  \bibfield  {author} {\bibinfo {author} {\bibfnamefont {Z.}~\bibnamefont
  {Jakubowski}}\ and\ \bibinfo {author} {\bibfnamefont {M.}~\bibnamefont
  {Kobel}},\ }\href@noop {} {\bibfield  {journal} {\bibinfo  {journal} {Nuclear
  Instruments and Methods in Physics Research Section A: Accelerators,
  Spectrometers, Detectors and Associated Equipment}\ }\textbf {\bibinfo
  {volume} {297}},\ \bibinfo {pages} {60} (\bibinfo {year} {1990})}\BibitemShut
  {NoStop}%
\bibitem [{\citenamefont {Bleicher}\ \emph {et~al.}(1999)\citenamefont
  {Bleicher} \emph {et~al.}}]{Bleicher:1999xi}%
  \BibitemOpen
  \bibfield  {author} {\bibinfo {author} {\bibfnamefont {M.}~\bibnamefont
  {Bleicher}} \emph {et~al.},\ }\href {\doibase 10.1088/0954-3899/25/9/308}
  {\bibfield  {journal} {\bibinfo  {journal} {J. Phys. G}\ }\textbf {\bibinfo
  {volume} {25}},\ \bibinfo {pages} {1859} (\bibinfo {year} {1999})},\ \Eprint
  {http://arxiv.org/abs/hep-ph/9909407} {arXiv:hep-ph/9909407} \BibitemShut
  {NoStop}%
\bibitem [{\citenamefont {Aab}\ \emph {et~al.}(2015)\citenamefont {Aab},
  \citenamefont {Abreu}, \citenamefont {Aglietta}, \citenamefont {Ahn},
  \citenamefont {Al~Samarai}, \citenamefont {Albuquerque}, \citenamefont
  {Allekotte}, \citenamefont {Allen}, \citenamefont {Allison}, \citenamefont
  {Almela} \emph {et~al.}}]{aab2015muons}%
  \BibitemOpen
  \bibfield  {author} {\bibinfo {author} {\bibfnamefont {A.}~\bibnamefont
  {Aab}}, \bibinfo {author} {\bibfnamefont {P.}~\bibnamefont {Abreu}}, \bibinfo
  {author} {\bibfnamefont {M.}~\bibnamefont {Aglietta}}, \bibinfo {author}
  {\bibfnamefont {E.}~\bibnamefont {Ahn}}, \bibinfo {author} {\bibfnamefont
  {I.}~\bibnamefont {Al~Samarai}}, \bibinfo {author} {\bibfnamefont
  {I.}~\bibnamefont {Albuquerque}}, \bibinfo {author} {\bibfnamefont
  {I.}~\bibnamefont {Allekotte}}, \bibinfo {author} {\bibfnamefont
  {J.}~\bibnamefont {Allen}}, \bibinfo {author} {\bibfnamefont
  {P.}~\bibnamefont {Allison}}, \bibinfo {author} {\bibfnamefont
  {A.}~\bibnamefont {Almela}},  \emph {et~al.},\ }\href@noop {} {\bibfield
  {journal} {\bibinfo  {journal} {Physical Review D}\ }\textbf {\bibinfo
  {volume} {91}},\ \bibinfo {pages} {032003} (\bibinfo {year}
  {2015})}\BibitemShut {NoStop}%
\bibitem [{\citenamefont {Aab}\ \emph {et~al.}(2021{\natexlab{c}})\citenamefont
  {Aab}, \citenamefont {Abreu}, \citenamefont {Aglietta}, \citenamefont
  {Albury}, \citenamefont {Allekotte}, \citenamefont {Almela}, \citenamefont
  {Alvarez-Mu{\~n}iz}, \citenamefont {Batista}, \citenamefont {Anastasi},
  \citenamefont {Anchordoqui} \emph {et~al.}}]{aab2021measurement}%
  \BibitemOpen
  \bibfield  {author} {\bibinfo {author} {\bibfnamefont {A.}~\bibnamefont
  {Aab}}, \bibinfo {author} {\bibfnamefont {P.}~\bibnamefont {Abreu}}, \bibinfo
  {author} {\bibfnamefont {M.}~\bibnamefont {Aglietta}}, \bibinfo {author}
  {\bibfnamefont {J.~M.}\ \bibnamefont {Albury}}, \bibinfo {author}
  {\bibfnamefont {I.}~\bibnamefont {Allekotte}}, \bibinfo {author}
  {\bibfnamefont {A.}~\bibnamefont {Almela}}, \bibinfo {author} {\bibfnamefont
  {J.}~\bibnamefont {Alvarez-Mu{\~n}iz}}, \bibinfo {author} {\bibfnamefont
  {R.~A.}\ \bibnamefont {Batista}}, \bibinfo {author} {\bibfnamefont {G.~A.}\
  \bibnamefont {Anastasi}}, \bibinfo {author} {\bibfnamefont {L.}~\bibnamefont
  {Anchordoqui}},  \emph {et~al.},\ }\href@noop {} {\bibfield  {journal}
  {\bibinfo  {journal} {Physical review letters}\ }\textbf {\bibinfo {volume}
  {126}},\ \bibinfo {pages} {152002} (\bibinfo {year}
  {2021}{\natexlab{c}})}\BibitemShut {NoStop}%
\bibitem [{\citenamefont {Albrecht}\ \emph {et~al.}(2022)\citenamefont
  {Albrecht}, \citenamefont {Cazon}, \citenamefont {Dembinski}, \citenamefont
  {Fedynitch}, \citenamefont {Kampert}, \citenamefont {Pierog}, \citenamefont
  {Rhode}, \citenamefont {Soldin}, \citenamefont {Spaan}, \citenamefont
  {Ulrich} \emph {et~al.}}]{albrecht2022muon}%
  \BibitemOpen
  \bibfield  {author} {\bibinfo {author} {\bibfnamefont {J.}~\bibnamefont
  {Albrecht}}, \bibinfo {author} {\bibfnamefont {L.}~\bibnamefont {Cazon}},
  \bibinfo {author} {\bibfnamefont {H.}~\bibnamefont {Dembinski}}, \bibinfo
  {author} {\bibfnamefont {A.}~\bibnamefont {Fedynitch}}, \bibinfo {author}
  {\bibfnamefont {K.-H.}\ \bibnamefont {Kampert}}, \bibinfo {author}
  {\bibfnamefont {T.}~\bibnamefont {Pierog}}, \bibinfo {author} {\bibfnamefont
  {W.}~\bibnamefont {Rhode}}, \bibinfo {author} {\bibfnamefont
  {D.}~\bibnamefont {Soldin}}, \bibinfo {author} {\bibfnamefont
  {B.}~\bibnamefont {Spaan}}, \bibinfo {author} {\bibfnamefont
  {R.}~\bibnamefont {Ulrich}},  \emph {et~al.},\ }\href@noop {} {\bibfield
  {journal} {\bibinfo  {journal} {Astrophysics and Space Science}\ }\textbf
  {\bibinfo {volume} {367}},\ \bibinfo {pages} {27} (\bibinfo {year}
  {2022})}\BibitemShut {NoStop}%
\bibitem [{\citenamefont {Kampert}\ and\ \citenamefont
  {Unger}(2012)}]{kampert2012measurements}%
  \BibitemOpen
  \bibfield  {author} {\bibinfo {author} {\bibfnamefont {K.-H.}\ \bibnamefont
  {Kampert}}\ and\ \bibinfo {author} {\bibfnamefont {M.}~\bibnamefont
  {Unger}},\ }\href@noop {} {\bibfield  {journal} {\bibinfo  {journal}
  {Astroparticle Physics}\ }\textbf {\bibinfo {volume} {35}},\ \bibinfo {pages}
  {660} (\bibinfo {year} {2012})}\BibitemShut {NoStop}%
\bibitem [{\citenamefont {Abreu}\ \emph {et~al.}(2012)\citenamefont {Abreu},
  \citenamefont {Aglietta}, \citenamefont {Ahn}, \citenamefont {Albuquerque},
  \citenamefont {Allard}, \citenamefont {Allekotte}, \citenamefont {Allen},
  \citenamefont {Allison}, \citenamefont {Almeda}, \citenamefont {Castillo}
  \emph {et~al.}}]{abreu2012measurement}%
  \BibitemOpen
  \bibfield  {author} {\bibinfo {author} {\bibfnamefont {P.}~\bibnamefont
  {Abreu}}, \bibinfo {author} {\bibfnamefont {M.}~\bibnamefont {Aglietta}},
  \bibinfo {author} {\bibfnamefont {E.}~\bibnamefont {Ahn}}, \bibinfo {author}
  {\bibfnamefont {I.~F. d.~M.}\ \bibnamefont {Albuquerque}}, \bibinfo {author}
  {\bibfnamefont {D.}~\bibnamefont {Allard}}, \bibinfo {author} {\bibfnamefont
  {I.}~\bibnamefont {Allekotte}}, \bibinfo {author} {\bibfnamefont
  {J.}~\bibnamefont {Allen}}, \bibinfo {author} {\bibfnamefont
  {P.}~\bibnamefont {Allison}}, \bibinfo {author} {\bibfnamefont
  {A.}~\bibnamefont {Almeda}}, \bibinfo {author} {\bibfnamefont {J.~A.}\
  \bibnamefont {Castillo}},  \emph {et~al.},\ }\href@noop {} {\bibfield
  {journal} {\bibinfo  {journal} {Physical Review Letters}\ }\textbf {\bibinfo
  {volume} {109}},\ \bibinfo {pages} {062002} (\bibinfo {year}
  {2012})}\BibitemShut {NoStop}%
\bibitem [{\citenamefont {Prado}(2019)}]{prado2019tests}%
  \BibitemOpen
  \bibfield  {author} {\bibinfo {author} {\bibfnamefont {R.~R.}\ \bibnamefont
  {Prado}},\ }in\ \href@noop {} {\emph {\bibinfo {booktitle} {EPJ Web of
  Conferences}}},\ Vol.\ \bibinfo {volume} {208}\ (\bibinfo {organization} {EDP
  Sciences},\ \bibinfo {year} {2019})\ p.\ \bibinfo {pages} {08003}\BibitemShut
  {NoStop}%
\bibitem [{\citenamefont {Cazon}\ \emph {et~al.}(2021)\citenamefont {Cazon},
  \citenamefont {Concei{\c{c}}{\~a}o}, \citenamefont {Martins},\ and\
  \citenamefont {Riehn}}]{cazon2021constraining}%
  \BibitemOpen
  \bibfield  {author} {\bibinfo {author} {\bibfnamefont {L.}~\bibnamefont
  {Cazon}}, \bibinfo {author} {\bibfnamefont {R.}~\bibnamefont
  {Concei{\c{c}}{\~a}o}}, \bibinfo {author} {\bibfnamefont {M.~A.}\
  \bibnamefont {Martins}}, \ and\ \bibinfo {author} {\bibfnamefont
  {F.}~\bibnamefont {Riehn}},\ }\href@noop {} {\bibfield  {journal} {\bibinfo
  {journal} {Physical Review D}\ }\textbf {\bibinfo {volume} {103}},\ \bibinfo
  {pages} {022001} (\bibinfo {year} {2021})}\BibitemShut {NoStop}%
\bibitem [{\citenamefont {Heitler}\ and\ \citenamefont
  {Ma}(1948)}]{heitler1948quantum}%
  \BibitemOpen
  \bibfield  {author} {\bibinfo {author} {\bibfnamefont {W.}~\bibnamefont
  {Heitler}}\ and\ \bibinfo {author} {\bibfnamefont {S.}~\bibnamefont {Ma}},\
  }in\ \href@noop {} {\emph {\bibinfo {booktitle} {Proceedings of the Royal
  Irish Academy. Section A: Mathematical and Physical Sciences}}},\
  Vol.~\bibinfo {volume} {52}\ (\bibinfo {organization} {JSTOR},\ \bibinfo
  {year} {1948})\ pp.\ \bibinfo {pages} {109--125}\BibitemShut {NoStop}%
\bibitem [{\citenamefont {Matthews}(2005)}]{matthews2005heitler}%
  \BibitemOpen
  \bibfield  {author} {\bibinfo {author} {\bibfnamefont {J.}~\bibnamefont
  {Matthews}},\ }\href@noop {} {\bibfield  {journal} {\bibinfo  {journal}
  {Astroparticle Physics}\ }\textbf {\bibinfo {volume} {22}},\ \bibinfo {pages}
  {387} (\bibinfo {year} {2005})}\BibitemShut {NoStop}%
\bibitem [{\citenamefont {Veberi{\v{c}}}(2012)}]{veberivc2012lambert}%
  \BibitemOpen
  \bibfield  {author} {\bibinfo {author} {\bibfnamefont {D.}~\bibnamefont
  {Veberi{\v{c}}}},\ }\href@noop {} {\bibfield  {journal} {\bibinfo  {journal}
  {Computer Physics Communications}\ }\textbf {\bibinfo {volume} {183}},\
  \bibinfo {pages} {2622} (\bibinfo {year} {2012})}\BibitemShut {NoStop}%
\bibitem [{\citenamefont {Kalmykov}\ \emph {et~al.}(2003)\citenamefont
  {Kalmykov}, \citenamefont {Alekseeva}, \citenamefont {Bergmann},
  \citenamefont {Chernatkin}, \citenamefont {Engel}, \citenamefont {Heck},
  \citenamefont {Moyon}, \citenamefont {Ostapchenko}, \citenamefont {Pierog},
  \citenamefont {Thouw} \emph {et~al.}}]{kalmykov2003one}%
  \BibitemOpen
  \bibfield  {author} {\bibinfo {author} {\bibfnamefont {N.}~\bibnamefont
  {Kalmykov}}, \bibinfo {author} {\bibfnamefont {M.}~\bibnamefont {Alekseeva}},
  \bibinfo {author} {\bibfnamefont {T.}~\bibnamefont {Bergmann}}, \bibinfo
  {author} {\bibfnamefont {V.}~\bibnamefont {Chernatkin}}, \bibinfo {author}
  {\bibfnamefont {R.}~\bibnamefont {Engel}}, \bibinfo {author} {\bibfnamefont
  {D.}~\bibnamefont {Heck}}, \bibinfo {author} {\bibfnamefont {J.}~\bibnamefont
  {Moyon}}, \bibinfo {author} {\bibfnamefont {S.}~\bibnamefont {Ostapchenko}},
  \bibinfo {author} {\bibfnamefont {T.}~\bibnamefont {Pierog}}, \bibinfo
  {author} {\bibfnamefont {T.}~\bibnamefont {Thouw}},  \emph {et~al.},\ }in\
  \href@noop {} {\emph {\bibinfo {booktitle} {International Cosmic Ray
  Conference}}},\ Vol.~\bibinfo {volume} {2}\ (\bibinfo {year} {2003})\ p.\
  \bibinfo {pages} {511}\BibitemShut {NoStop}%
\bibitem [{\citenamefont {Cazon}(2019)}]{cazon2019probing}%
  \BibitemOpen
  \bibfield  {author} {\bibinfo {author} {\bibfnamefont {L.}~\bibnamefont
  {Cazon}},\ }\href@noop {} {\bibfield  {journal} {\bibinfo  {journal} {arXiv
  preprint arXiv:1909.02962}\ } (\bibinfo {year} {2019})}\BibitemShut {NoStop}%
\bibitem [{\citenamefont {G{\'e}ron}(2022)}]{geron2022hands}%
  \BibitemOpen
  \bibfield  {author} {\bibinfo {author} {\bibfnamefont {A.}~\bibnamefont
  {G{\'e}ron}},\ }\href@noop {} {\emph {\bibinfo {title} {Hands-on machine
  learning with {S}cikit-{L}earn, {K}eras, and {T}ensor{F}low}}}\ (\bibinfo
  {publisher} {" O'Reilly Media, Inc."},\ \bibinfo {year} {2022})\BibitemShut
  {NoStop}%
\bibitem [{\citenamefont {Glorot}\ and\ \citenamefont
  {Bengio}(2010)}]{glorot2010understanding}%
  \BibitemOpen
  \bibfield  {author} {\bibinfo {author} {\bibfnamefont {X.}~\bibnamefont
  {Glorot}}\ and\ \bibinfo {author} {\bibfnamefont {Y.}~\bibnamefont
  {Bengio}},\ }in\ \href@noop {} {\emph {\bibinfo {booktitle} {Proceedings of
  the thirteenth international conference on artificial intelligence and
  statistics}}}\ (\bibinfo {organization} {JMLR Workshop and Conference
  Proceedings},\ \bibinfo {year} {2010})\ pp.\ \bibinfo {pages}
  {249--256}\BibitemShut {NoStop}%
\bibitem [{\citenamefont {Goos}(2023)}]{Goos:2022dqv}%
  \BibitemOpen
  \bibfield  {author} {\bibinfo {author} {\bibfnamefont {I.~A.}\ \bibnamefont
  {Goos}},\ }\emph {\bibinfo {title} {{Determination of physical properties of
  high-energy hadronic interactions from the $X_{max}-N_\mu$
  anticorrelation}}},\ \href {\doibase 10.5445/IR/1000156068} {Ph.D. thesis},\
  \bibinfo  {school} {KIT, Karlsruhe, Dept. Phys.} (\bibinfo {year}
  {2023})\BibitemShut {NoStop}%
\bibitem [{\citenamefont {Lista}(2017)}]{lista2017statistical}%
  \BibitemOpen
  \bibfield  {author} {\bibinfo {author} {\bibfnamefont {L.}~\bibnamefont
  {Lista}},\ }\href@noop {} {\emph {\bibinfo {title} {Statistical methods for
  data analysis in particle physics}}},\ Vol.\ \bibinfo {volume} {941}\
  (\bibinfo  {publisher} {Springer},\ \bibinfo {year} {2017})\BibitemShut
  {NoStop}%
\end{thebibliography}%

\end{document}